\documentclass[runningheads]{llncs}

\usepackage{amsmath}
\usepackage{amsfonts}
\usepackage{amssymb}
\usepackage{booktabs}
\usepackage{color}
\usepackage{graphicx}
\usepackage[colorlinks]{hyperref}
\usepackage[switch]{lineno}
\usepackage[dvipsnames]{xcolor}
\usepackage{subcaption}  % subfigures
\usepackage{tikz}        % drawing trees
\usepackage{forest}      % drawing trees
\usepackage{wrapfig}

\graphicspath{{figures/}}

\tikzset{
  solid node/.style={circle,draw,inner sep=1.2,fill=black},
  hollow node/.style={circle,draw,inner sep=1.2},
  no node/.style={inner sep=1.2},
  left label/.style={above left,midway},
  right label/.style={above right,midway},
  mid label/.style={above,midway}
}
\newcommand*{\bigdot}{\raisebox{-0.25ex}{\scalebox{2.0}{$\cdot$}}}

\newcommand{\ignore}[1]{}

\title{Ballot-Polling Audits of Instant-Runoff Voting Elections with a
Dirichlet-Tree Model\thanks{Published in: ESORICS 2022 Workshops, EIS 2022.
LNCS 13785, pp.\ 525--540, Springer, Cham (2023).
\url{https://doi.org/10.1007/978-3-031-25460-4_30}}}

\titlerunning{Ballot-Polling Audits of IRV Elections with a Dirichlet-Tree
Model}

\author{
Floyd Everest     \inst{1}   \orcidID{0000-0002-2726-6736}  \and
Michelle Blom     \inst{2}   \orcidID{0000-0002-0459-9917}  \and
Philip B. Stark   \inst{3}   \orcidID{0000-0002-3771-9604}  \and
Peter J. Stuckey  \inst{4}   \orcidID{0000-0003-2186-0459}  \and
Vanessa Teague    \inst{5,6} \orcidID{0000-0003-2648-2565}  \and
Damjan Vukcevic   \inst{1,7} \orcidID{0000-0001-7780-9586}}

\authorrunning{Everest, Blom, Stark, Stuckey, Teague, Vukcevic}
\institute{
School of Mathematics and Statistics, University of Melbourne, Parkville,
Australia
\and
School of Computing and Information Systems, University of Melbourne,
Parkville, Australia
\and
Department of Statistics, University of California, Berkeley, CA, USA
\and
Department of Data Science and AI, Monash University, Clayton, Australia
\and
Thinking Cybersecurity Pty.\ Ltd., Melbourne, Australia
\and
The Australian National University, Canberra, Australia
\and
Melbourne Integrative Genomics, University of Melbourne, Parkville,
Australia \\
\email{damjan.vukcevic@unimelb.edu.au}}

\begin{document}

\maketitle

\begin{abstract}
Instant-runoff voting (IRV) is used in several countries around the world.  It
requires voters to rank candidates in order of preference, and uses a counting
algorithm that is more complex than systems such as first-past-the-post or
scoring rules.  An even more complex system, the single transferable vote
(STV), is used when multiple candidates need to be elected.  The complexity of
these systems has made it difficult to audit the election outcomes.  There is
currently no known risk-limiting audit (RLA) method for STV, other than a full
manual count of the ballots.

A new approach to auditing these systems was recently proposed, based on a
Dirichlet-tree model.  We present a detailed analysis of this approach for
ballot-polling Bayesian audits of IRV elections.  We compared several choices
for the prior distribution, including some approaches using a Bayesian
bootstrap (equivalent to an improper prior).  Our findings include that the
bootstrap-based approaches can be adapted to perform similarly to a full
Bayesian model in practice, and that an overly informative prior can give
counter-intuitive results.  Via carefully chosen examples, we show why creating
an RLA with this model is challenging, but we also suggest ways to overcome
this.

As well as providing a practical and computationally feasible implementation of
a Bayesian IRV audit, our work is important in laying the foundation for an RLA
for STV elections.
\end{abstract}

% ---------------------------------------------------------------------------

\section{Introduction}

Audits of elections should provide rigorous statistical evidence in favour of
the reported outcomes, or otherwise correct the result if the outcome is wrong.
When the reported electoral outcome is correct, statistical audits can usually
do so with less effort than a full manual count of all the ballots.

A \emph{risk-limiting audit} (RLA) guarantees that if the reported outcome is
wrong, there is a large chance that the audit will correct it; while if the
reported outcome is correct, the audit does not change it \cite{stark08a}.  The
risk limit is the maximum chance that a wrong outcome will not be corrected by
the audit.  RLAs have been developed for a wide variety of voting systems,
including plurality, multi-winner plurality, supermajority, STAR-Voting, and
proportional representation schemes \cite{shangrla,StarkTeague2014,voting21}.
Indeed, any social choice function (method for determining the winner) that
asks voters to select one or more candidates on their ballot paper, or to
assign `scores' to them, can be audited with existing methods \cite{shangrla}.
However, some jurisdictions use more complex social choice functions.

\emph{Ranked-choice voting} requires voters to rank candidates (or political
parties parties, or other groupings of candidates) in order of preference.
Some elections require a complete ranking, while others allow partial rankings.
The way the votes are counted can be quite involved.  A commonly used system is
\emph{instant-runoff voting} (IRV), which involves tallying the
first-preference counts and then iteratively eliminating the candidate with the
lowest tally and redistributing their ballots (according to the next preference
on their ballots) until one candidate achieves a majority of votes.  There are
efficient risk-limiting methods to audit IRV contests
\cite{blom2019raire,shangrla,blomEtal20}.

An even more complex system, for which no risk-limiting method is known, is the
\emph{single transferable vote} (STV), designed to elect several candidates;
its counting algorithm involves transferring `surplus' votes from one candidate
to others in proportion to how much that candidate's tally exceeds a `quota'.

There are two ways in which ranked-choice voting is more complex than many
other systems.  First, the number of ways that a voter can fill out their
ballot is very large (there are $k!$ ways to rank $k$ candidates).  This makes
auditing mathematically challenging: the statistical inference required is now
in a very high-dimensional parameter space.  Second, the social choice
functions are typically combinatorially complex, making it challenging even to
calculate the `margin' (see \autoref{sec:margin}).

SHANGRLA \cite{shangrla} is a general framework for RLAs that covers a wide
variety of audit types (such as ballot-polling, comparison, stratified) and
voting systems (including the various systems mentioned earlier).

Some RLAs have already been developed for ranked voting:
RAIRE~\cite{blom2019raire} for IRV elections, and a similar recent method for
2-seat STV elections~\cite{blom2021stv}.  Both methods address the high
dimensionality by projecting into lower dimensions.  This allows them to use
the SHANGRLA framework, which also comes with the benefit of automatically
allowing different types of audits (ballot-polling, comparison, etc.).  For
most social choice functions for which there are currently SHANGRLA audits,
SHANGRLA tests conditions (`assertions') that are necessary and sufficient for
the outcome to be correct.  In contrast, for IRV and STV, the conditions are
\emph{sufficient} not always \emph{necessary}.  In particular, the projections
used in \cite{blom2019raire,blom2021stv} check some but not all of the possible
elimination sequences that lead to the reported winner(s) really winning.   It
is possible that the reported winner(s) really won, but through a different
elimination sequence---in which case the audit could lead to an unnecessary
full count.  Several open problems thus remain, including a feasible RLA for
STV elections of more than 2 seats, and an RLA for IRV when individual vote
records are not available.

One strategy is to tackle the statistical inference problem directly in its
natural (high-dimensional) space.  Bayesian audits \cite{rivest2012,rivest2018}
are typically set up in this way, by specifying a model and prior distribution
over this space.  A naive application of Bayesian inference to IRV or STV will
fail when there are more than a handful of candidates, because the dimension of
the parameter space will be too high: the models behave poorly and the
computational burden is prohibitive.  An early Bayesian approach to auditing of
STV \cite{chilingirian2016auditing} avoided specifying a full model, using a
(Bayesian) bootstrap approach instead.  This approximates a full Bayesian
inference but is less computationally demanding.

We recently proposed a full Bayesian method using a Dirichlet-tree model to
allow inference in high dimensions in a computationally feasible manner, and
provided a proof-of-concept \cite{everest2022evote}.
Here, we provide a thorough description and an extensive analysis of the
approach using data from a diverse set of real elections.  We do so for
\emph{ballot-polling} audits of IRV elections.  Ballot-polling audits manually
interpret the votes on randomly selected ballots, but do not compare those
interpretations to how the voting system interpreted the same ballots.  They
are the simplest method to implement because they do not require much data from
the voting system (but they generally require larger audit samples than
\emph{comparison audits}, which compare human interpretations of the votes to
the system's record of the votes, either for individual ballots or clusters of
ballots).

The Dirichlet-tree model is very flexible in how it can be set up.  It also
unifies many previous proposals, which turn out to be special cases of the
model.  We evaluate these and show that many of them work well in practice.
Furthermore, we illustrate some unexpected behaviour if the model is set up in
specific ways.

One important consideration in practice is whether this type of audit can be
made risk-limiting.  It has been shown that Bayesian audits for simpler
elections can be calibrated to limit risk \cite{huang_unified_2020}, however
whether this is possible for IRV and STV is still an open problem.  We use an
artifical example to illustrate why this is likely to be challenging, and also
point to some promising new techniques that could help solve this problem.

Our work provides a practical Bayesian IRV audit, with key insights and clear
scope for how it can be generalised to STV or other ranked voting systems.  It
also lays the foundation for an RLA for such systems, with explicit suggestions
for how the method could be adapted to limit risk.

% ---------------------------------------------------------------------------

\section{Methods}

\subsection{Election audits and the Dirichlet-tree model}

We adopt and expand on our previous framework and notation
\cite{everest2022evote}.
%We describe the main concepts at a relatively high level.  Further
%mathematical detail can be found in standard textbooks on Bayesian inference.

Suppose there are $k$ candidates in a given contest, and $K$ ways for voters to
vote.
%\footnote{%
%If we require voters to specify a complete ranking of the candidates,
%then $K = k!$.}
Across the whole contest, let $p_i$ be the proportion of ballots that are of
type $i$, for $i = 1, 2, \dots, K$.  The election outcome is completely
determined by $\vec{p} = (p_1, p_2, \dots, p_K)$.  If we knew $\vec{p}$, we
could verify the reported outcome.

An election audit involves sampling ballots from the contest and using a
statistical model to infer the true outcome (e.g.\ by estimating $\vec{p}$), to
some desired level of certainty.
In a sample of ballots, let $n_i$ be the number of observed ballots of type
$i$.  If sampling the ballots at random with replacement, the distribution of
the tallies $\vec{n} = (n_1, n_2, \dots, n_K)$ is given by a multinomial
distribution.

This distribution describes the possible variation in the data.  In a Bayesian
audit, we combine this together with a \emph{prior} distribution, which
describes our uncertainty about the election outcome by specifying a
distribution of possible values for $\vec{p}$.  Given the data, this
distribution gets `updated' (using Bayes' theorem) to a \emph{posterior}
distribution: also a distribution for $\vec{p}$, but with the probability mass
shifted to values that are more consistent with the data.

A distribution for $\vec{p}$ induces a distribution for the winning
candidate(s).  That is, for each candidate, we obtain a probability that they
won the election.
%\footnote{%
%We also refer to these as `posterior' probabilities.  It is simply the
%distribution for $\vec{p}$ projected into a new space.}
These are used to determine whether the data provide sufficient evidence in
favour of the reported outcome.

A standard process for the audit would involve sampling some ballots,
calculating the posterior probability for the reported winner(s), and comparing
this to a desired threshold value (e.g.\ 99\%).  If the posterior exceeds the
threshold, we terminate the audit.  Otherwise, we sample more ballots and
repeat the process.  We stop either when we exceed the threshold, or have
sampled all ballots, or possibly reached a specified sampling limit (e.g.\
imposed for cost reasons).

Bayesian inference is internally consistent irrespective of when we terminate
the sampling, if the prior exactly reflects the analyst's beliefs.  However,
the chance the audit stops short of a full count when the outcome is wrong can
depend on whether the prior is generative, i.e.\ how Nature generates votes
\cite{deheide2021}.

\subsubsection{Dirichlet and Dirichlet-tree priors.}

To start the process, we need to choose a prior distribution.  When using a
multinomial model for the data, a popular choice for the prior is a Dirichlet
distribution.  This is defined by $\vec{a} = (a_1, a_2, \dots, a_K)$, where
$a_i$ is called the \emph{concentration parameter}, or simply the
\emph{weight}, for ballot type $i$.  Larger values of $a_i$ lead to $p_i$ being
more likely to have higher values (the distribution is more `concentrated'
around that ballot type).  The special case $a_1 = a_2 = \dots = a_K = 1$ gives
a uniform distribution for $\vec{p}$.

The Dirichlet distribution is a \emph{conjugate} prior (the property that if
the prior is Dirichlet, then the posterior will also be Dirichlet) and is easy
to compute from the data: the posterior will have weights of the form $a_i +
n_i$.

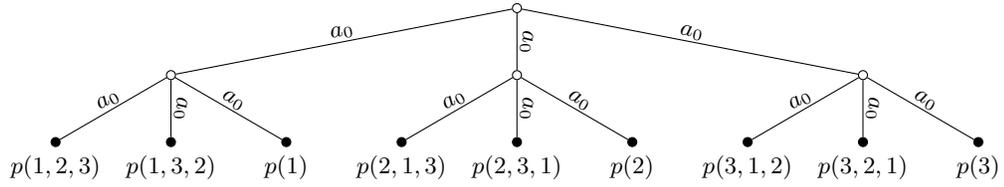
\begin{figure}[t]
\centering
\begin{forest}
    for tree={
        s sep=1.4cm,
        l sep=20pt,
    },
    a0/.style={
      edge label={node [midway,inner sep=1pt,above,sloped,font=\footnotesize] {$a_0#1$}},
    },
[, hollow node
  [, hollow node, a0
    [, solid node, label={below:$p(1,2,3)$}, a0]
    [, solid node, label={below:$p(1,3,2)$}, a0]
    [, solid node, label={below:$p(1)$}, a0]
  ]
  [, hollow node, a0
    [, solid node, label={below:$p(2,1,3)$}, a0]
    [, solid node, label={below:$p(2,3,1)$}, a0]
    [, solid node, label={below:$p(2)$}, a0]
  ]
  [, hollow node, a0
    [, solid node, label={below:$p(3,1,2)$}, a0]
    [, solid node, label={below:$p(3,2,1)$}, a0]
    [, solid node, label={below:$p(3)$}, a0]
  ]
]
\end{forest}
\caption{Dirichlet-tree prior for IRV ballots with 3 candidates, with a weight
of $a_0$ in each branch.  For simplicity, a redundant bottom layer is not
included in the tree, since ballots of the form $(1,2,3)$ are equivalent to
$(1,2)$.}
\label{fig:irv_tree_prior}
\end{figure}

As we have previously described \cite{everest2022evote}, the Dirichlet becomes
unwieldy as $K$ grows very large, and an alternative called a Dirichlet-tree
was proposed \cite{dirichlet-tree-distribution,dennis1991hyper}.  This arranges
the ballot types into a tree structure, with the branches in the tree
describing choices of candidates for each place in the preference ordering.  A
path through the tree corresponds to a particular ordering of the candidates,
and hence a ballot type.  We can also accommodate partial orderings by
including `termination' branches in the tree.  See \autoref{fig:irv_tree_prior}
for an example.

To complete the specification, we place a Dirichlet distribution at each node
in the tree, to model the conditional split of preferences locally at that
node.  This set of nested distributions together gives a complete distribution
across all ballot types.  We refer to this as a \emph{Dirichlet-tree
distribution}, with the parameters being the weights in each branch of the
tree.  The distribution turns out to also be conjugate and can be updated
efficiently: the posterior is a Dirichlet-tree with weights of the form $a_j +
n_j$ for branch $j$, where $n_j$ is the number of ballots in the sample whose
path through the tree traverses that branch.

Finally, we need to specify the (prior) weights for each branch.  We
investigated several choices for this, see \autoref{sec:choice-of-prior}.

\subsubsection{Dirichlet equivalence.}

\begin{figure}
\centering
\begin{subfigure}[b]{0.45\textwidth}
\centering
\begin{forest}
    for tree={
        s sep=1in,
        l sep=10pt,
        inner sep=.5pt,
    },
    a0/.style={
      edge label={node [midway,inner sep=1pt,above,sloped,font=\footnotesize] {$#1$}},
    },
[, hollow node
    [, hollow node, a0={\alpha_1+\alpha_2}
        [, solid node, label={below:$p_1$}, a0={\alpha_1}]
        [, solid node, label={below:$p_2$}, a0={\alpha_2}]
    ]
    [, solid node, label={below:$p_3$}, a0={\alpha_3}]
]
\end{forest}
\end{subfigure}
\hfill
\begin{subfigure}[b]{0.45\textwidth}
\centering
\begin{forest}
    for tree={
        s sep=1in,
        l sep=20pt,
    },
    a0/.style={
      edge label={node [midway,inner sep=1pt,above,sloped,font=\footnotesize] {$#1$}},
    },
[, hollow node
    [, solid node, label={below:$p_1$}, a0={\alpha_1}]
    [, solid node, label={below:$p_2$}, a0={\alpha_2}]
    [, solid node, label={below:$p_3$}, a0={\alpha_3}]
]
\end{forest}
\end{subfigure}
\caption{Two equivalent Dirichlet-trees. The one on the right is explicitly
just a Dirichlet distribution over the 3 categories.}
\label{fig:dtree-dir-equiv}
\end{figure}
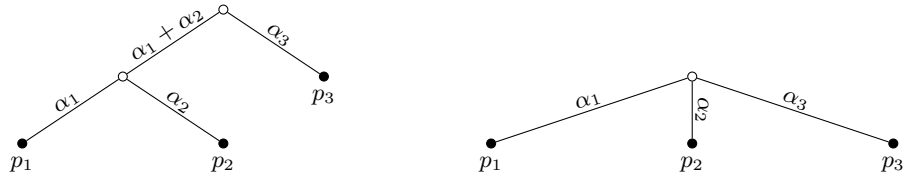

The Dirichlet-tree generalises the Dirichlet.  Specifically, any Dirichlet-tree
for which the weight in each node's parent branch is equal to the sum of the
weights in all of its child branches, is equivalent to a Dirichlet distribution
where we remove all of the internal nodes.  \autoref{fig:dtree-dir-equiv} shows
an example.  This equivalence allows us to use a single software implementation
to explore both Dirichlet and Dirichlet-tree priors.

\subsubsection{Implementing Dirichlet-tree audits for IRV elections.}

On a single IRV ballot, up to $k$ candidates are ordered from highest
preference to lowest.  The set of possible ballot types cast in an IRV election
are exactly the set of permutations on non-empty subsets of the $k$ candidates,
of which there are $f(k) = \sum_{i = 1}^k  \binom{k}{i}
i!$.\footnote{\url{https://oeis.org/A007526}} Note that $f(k)$ grows \emph{very
quickly} with $k$, e.g.\ $f(8) = 190,600$ while $f(18) = 1.7 \times 10^{16}$.
%In some IRV elections we can only rank at most $l$ candidates, in which case
%each ballot is a permutation of non-empty subsets of the $k$ candidates of
%size up to $l$, of which there are $f(k,l) = \sum_{i = 1}^l  \binom{k}{i} i!$.
To cope with this very large space, we use a Dirichlet-tree prior.

A naive approach to representing a Dirichlet-tree that assigns a non-zero
probability to each possible ballot type will struggle as $k$ gets large.  In
our experiments, we considered elections with $k$ as large as $18$.
Representing all tree nodes explicitly would require significant memory. We can
avoid this by only representing nodes that have been updated and now differ
from their default prior weight. This limits the memory required to $O(kn)$
where $n$ is the number of ballots sampled.

\begin{figure}
    \centering
    \begin{forest}
    for tree={
        s sep=5pt,
        l sep=45pt,
        inner sep=.5pt,
    },
    pruned/.style={
        edge=dotted,
    },
    perm/.style={
        content={$\begin{pmatrix}#1\end{pmatrix}$},
        align=center,
        tier=leaf,
    },
[, hollow node
    [, hollow node, tikz={\node [draw,red,fit to=tree, align=left, text height=10pt] {$B_1$};}
        [, hollow node
            [, perm={1\\2\\3\\4}, pruned]
            [, perm={1\\2\\4\\3}, pruned]
            [, perm={1\\2\\\bigdot\\\bigdot}]
        ]
        [, hollow node
            [, perm={1\\3\\2\\4}, pruned]
            [, perm={1\\3\\4\\2}, pruned]
            [, perm={1\\3\\\bigdot\\\bigdot}]
        ]
        [, hollow node
            [, perm={1\\4\\2\\3}, pruned]
            [, perm={1\\4\\3\\2}, pruned]
            [, perm={1\\4\\\bigdot\\\bigdot}]
        ]
        [, no node, pruned [, perm={1\\\bigdot\\\bigdot\\\bigdot}, pruned]]
    ]
    [$B_2$ [,phantom [,phantom [, phantom]]][,phantom][,phantom][,phantom][,phantom]]
    [$B_3$]
    [$B_4$ [,phantom][,phantom][,phantom][,phantom][,phantom][,phantom]]
    [, no node, pruned[, no node, pruned[, perm={\bigdot\\\bigdot\\\bigdot\\\bigdot}, pruned]]]
]
    \end{forest}
\caption{A Dirichlet-tree representing IRV ballots with minimum and maximum
depth set to 2 (thus, only allowing ballots that specify exactly two
candidates).  The dotted branches represent ballot types that have been
pruned.}
\label{fig:pirv_dtree_prior}
\end{figure}
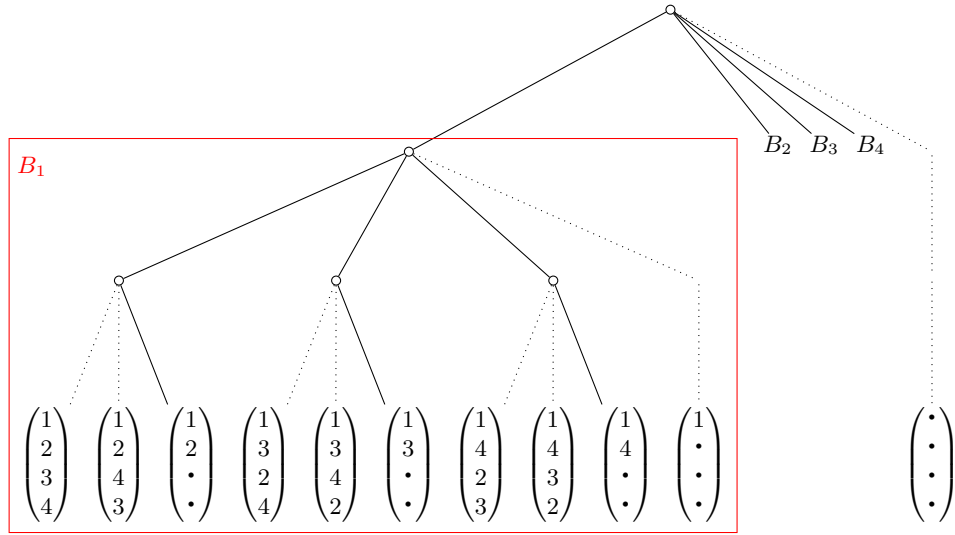

Our software implementation\footnote{%
Available at: \url{https://github.com/fleverest/elections.dtree}}
allows setting a minimum and maximum depth for the tree, which is equivalent to
requiring a min or max number of candidates be marked on the ballot.  See
\autoref{fig:pirv_dtree_prior} for an illustration.  In our experiments, we
always set a minimum depth of 1 (to rule out empty ballots) and sometimes also
set a maximum depth (depending on the validity criteria in each contest).

\subsection{Choice of prior distribution}
\label{sec:choice-of-prior}

\begin{description}
\item[Dirichlet-tree with equally weighted branches (EWB).]
This is parameterised by $a_0$, the prior weight for each branch.  As a
default, we used $a_0 = 1$, which is equivalent to a uniform distribution on
the probabilities at each node in the tree.  We also explored smaller and
larger values, up to $a_0 = 1000$.

\item[Bayesian bootstrap.]
We can make the prior less informative by reducing the weight of each branch.
In the limit where each weight is reduced to zero (equivalent to EWB with $a_0
= 0$), we obtain an improper prior.  The posterior will only have positive
probability (`support') on ballots that appear in the sample.  This model is
known as a Bayesian bootstrap \cite{rubin1981}.
%and can be considered as an approximation to a full Bayesian analysis.

\item[Bayesian bootstrap seeded with $b_0$ single-preference ballots
for each candidate.]
This prior was suggested by \cite{chilingirian2016auditing}.  If there are $k$
candidates, then we add $k \times b_0$ ballots in total to the tree.  The
special case $b_0 = 0$ recovers the Bayesian bootstrap.  We explored values up
to $b_0 = 1000$.

\item[Bayesian bootstrap with a minimum sample size.]
In other words, a Bayesian bootstrap with the requirement that we need to
sample at least some pre-specified number of ballots before we may stop the
audit and certify the election.
%\footnote{%
%This isn't actually a different prior distribution, but a slight modification
%to the decision procedure.  It is a simple mechanism to prevent early stopping
%due to the posterior being overly confident when given very little data.  It
%has a similar effect to making the prior more informative.}
%In the Bayesian bootstrap, the posterior is essentially an approximation that
%is based on the data only, and does not benefit from any regularisation that a
%more typical choice of prior would induce.  Hence it behaves poorly when there
%is very little data.
We set a minimum sample size of 20 when using this scheme.

\item[Dirichlet.]
%This is a special case of a Dirichlet-tree, with the restriction that the
%weights in each branch are equal to the sum of the weights in the immediate
%`downstream' branches.
We used Dirichlet distributions that gave all of the maximally specified ballot
types (up to the maximum depth allowed) an equal weight, $a_0$.  In the
tree-representation, these are the weights given to all lowest level branches,
with the other branches' weights determined by summing them up the tree.  Note
that the weights in the top-level branches will vary substantially based on the
structure of the tree, which depends on the number of candidates as well as the
choice of minimum and maximum depth.  We explored values of $a_0$ from 0.001 to
10.
\end{description}

\subsection{Data}

We used ballot data from elections in Australia and the USA.\footnote{%
All data were sourced from: \url{https://github.com/michelleblom/margin-irv}}
The Australian data included 93 contests all coming from the NSW 2015 lower
house election; each had 5--8 candidates and about 40k--50k ballots.
The USA data included 14 contests from elections in California and Colorado;
they were much more variable, having 4--18 candidates and the number of ballots
ranged from 2544 (Aspen) to 312,771 (Pierce).
In addition, we constructed 3 artificial `pathological' contests that were
specifically designed to be difficult to audit; each had 10 candidates and
11,000 ballots (see \autoref{sec:pathological} for more details).
The contests from California only allowed voters to mark their top 3
preferences on their ballots.  We encoded this in our models by restricting the
trees to have a maximum depth of 3.  We did the same for the pathological
contests.

\subsubsection{Margin.}
\label{sec:margin}

To provide context for interpreting the performance of the auditing methods, we
quantified how close each contest was by calculating the \emph{margin} using
\texttt{margin-irv}~\cite{blom2018margin}.
A positive margin is the minimum number of ballots that need to be changed in
order for the reported winner to \emph{no longer be the true winner}.
A negative margin is the minimum number (expressed as a negative integer) of
ballots that need to be changed in order for the reported winner to \emph{tie
with the true winner}.

\subsection{Benchmarking experiments}

To demonstrate our model, and compare the proposed choices for the prior
distribution, we simulated audits using the data for all of the contests
described above.  We used a procedure similar to earlier work
\cite{everest2022evote}.

\subsubsection{Main analyses.}

Our main analyses used the data faithfully, i.e.\ without adding any errors.
For each contest, we randomly shuffled the ballots 100 times, to simulate 100
different orderings.  Each simulated audit draws a sample of ballots one at a
time from a given ordering.  We used the same orderings (for each contest)
across all of the different auditing methods (e.g.\ different priors) that we
evaluated.

After each ballot is sampled, one at a time, we estimated the posterior
probabilities (for each candidate winning) by taking the mean of 100 Monte
Carlo draws from the posterior.  Specifically, we took draws from the posterior
predictive distribution of the ballot tallies across the whole contest (i.e.\
the full set of ballots), and applied the IRV social choice function to each;
this gave us Monte Carlo draws from the posterior on the winning candidate.

\begin{figure}
\centering
\includegraphics[width=\textwidth]{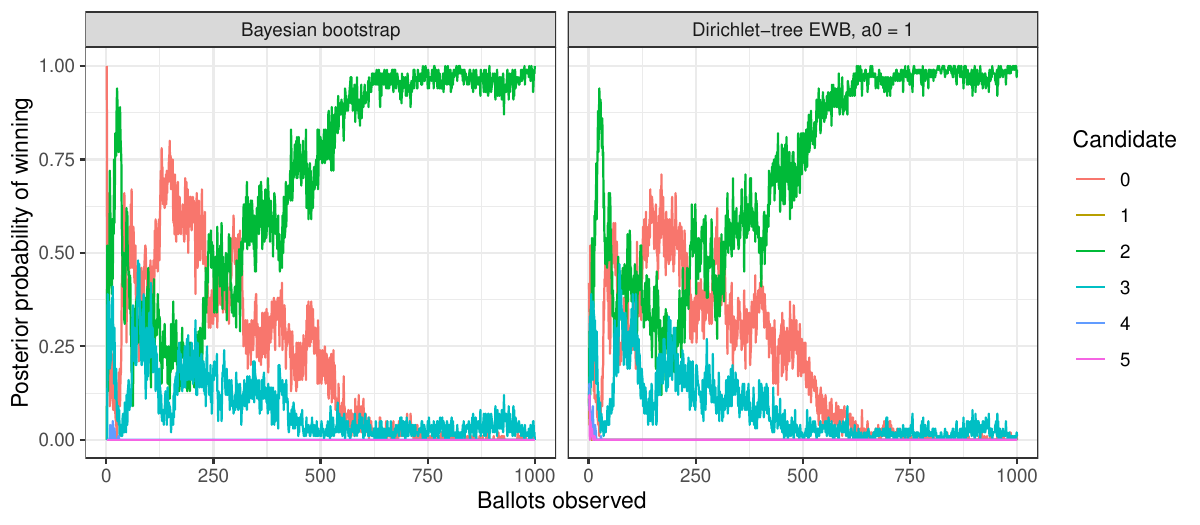}
\caption{\textbf{Posterior probabilities from an example audit.}
Posterior probabilities for each candidate, vs sample size, for a simulated
audit of the Lismore (NSW 2015) election, using two different priors (as
labelled).  The true winner is candidate 2 (green line).}
\label{fig:posterior-trace}
\end{figure}

We allowed samples of up to 1000 ballots.
\autoref{fig:posterior-trace} shows an example.

The possible outcomes for each audit are one of:
\begin{description}
\item[Certify.]  A desired threshold for the posterior probability is exceeded
    at some point during the sampling.  The audit terminates at this point, and
    we record the sample size (number of ballots sampled).  In our experiments,
    we used a threshold of 99\%.
\item[Do not certify.]  The desired threshold is not exceeded during the first
    1000 ballots.  (In practice, this could be a point at which the sampling is
    terminated and a full manual count conducted instead.)
\end{description}
Across the 100 orderings we can then calculate the \emph{certification rate}
(the proportion of the orderings that led to certification) and also the
\emph{mean sample size} (irrespective of certification).  There is some
redundancy in these measures: a low certification rate would usually lead to
the mean sample size being close to 1000.

This way of measuring the mean sample size is convenient in our context,
allowing easy comparison across different contests.  It only measures the
`work' required in the sampling part of the audit, and not in any potential
full manual count.  The `cost' of the latter could vary substantially across
contests, depending on how many ballots they involve.

Ideally, we would let the simulation run beyond 1000 ballots and simply record
the sample size at termination.  This would have required an impractical amount
of computation, given the number of experiments we ran.  Our choice to limit
each audit to 1000 ballots, and also to take only 100 draws from the posterior
for each calculation (see above), were purely pragmatic.  Nevertheless, they
were sufficient to explore the general behaviour of these models.

Note that in using 100 draws, our posterior probability estimates only had a
resolution of 0.01 and still have some Monte Carlo error (this is visible as
low-level `noise' in the curves in \autoref{fig:posterior-trace}).  If these
auditing methods were to be used in a real audit, we would only do such
calculations a few times and these constraints would be unnecessary.  It would
be straightfoward and computationally feasible to use sufficiently many draws
to eliminate the Monte Carlo error.

\subsubsection{Generating errors by permuting candidate labels.}
\label{sec:permuting-candidates}

The main analyses (described above) explore the case where there are no
tabulation errors.  In practice we would expect errors due to misinterpreting
marks on a ballot or other sporadic mistakes, or systemic issues in how ballots
are handled or interpreted.

There are many ways to model and explore the effect of such errors.  For
convenience, we used the following `trick' to rapidly simulate a large number
of contests where the reported outcome differed to the true outcome.

For any given contest, we can take any candidate to be the reported winner and
run the audit based on that candidate.  The calculations required to obtain the
posterior probabilities are the same regardless, only the stopping criterion
differs, and this is very fast to evaluate given a set of posterior
probabilities.  Thus, for a $k$-candidate contest, we obtain experimental
results for $k$ scenarios: one for which the reported winner is the true
winner, and $k-1$ scenarios where the reported winner is incorrect.  The
computational cost of the full $k$ scenarios is essentially the same as just a
single scenario.

We can interpret these scenarios in at least two different ways.  First, they
are equivalent to an error model where the candidate labels are switched (which
is conceivable in practice, e.g. via a software error, or deliberate
manipulation).

Alternatively, we can take them as an arbitrary set of examples where the
reported winner is incorrect.  This follows because the only information used
by these Bayesian audits is the reported winner; they do not make use of any
cast vote records.  Thus, whether the error rate is large or small is
immaterial, the same scenario could have been produced by two different
contests with substantially different error rates.  The only factors that will
affect performance are whether the reported winner is the true winner, and how
`close' the election is.  We measure the latter by calculating the
margin~\cite{blom2018margin}.

Across all of the contests, this permutation scheme generated a large and
diverse set of scenarios.

% ---------------------------------------------------------------------------

\section{Results}

\subsection{Comparing the priors}

We generated 100 random orderings of ballots for each of the 107 contests.  For
each such ordering of each contest, we simulated an audit using each of the
different priors, across several choices of parameter value (where applicable).
Each audit was allowed to sample up to 1000 ballots, terminating once the
posterior probability for the reported winner exceeded 99\%.

\begin{figure}
\centering
\includegraphics[width=\textwidth,page=2]{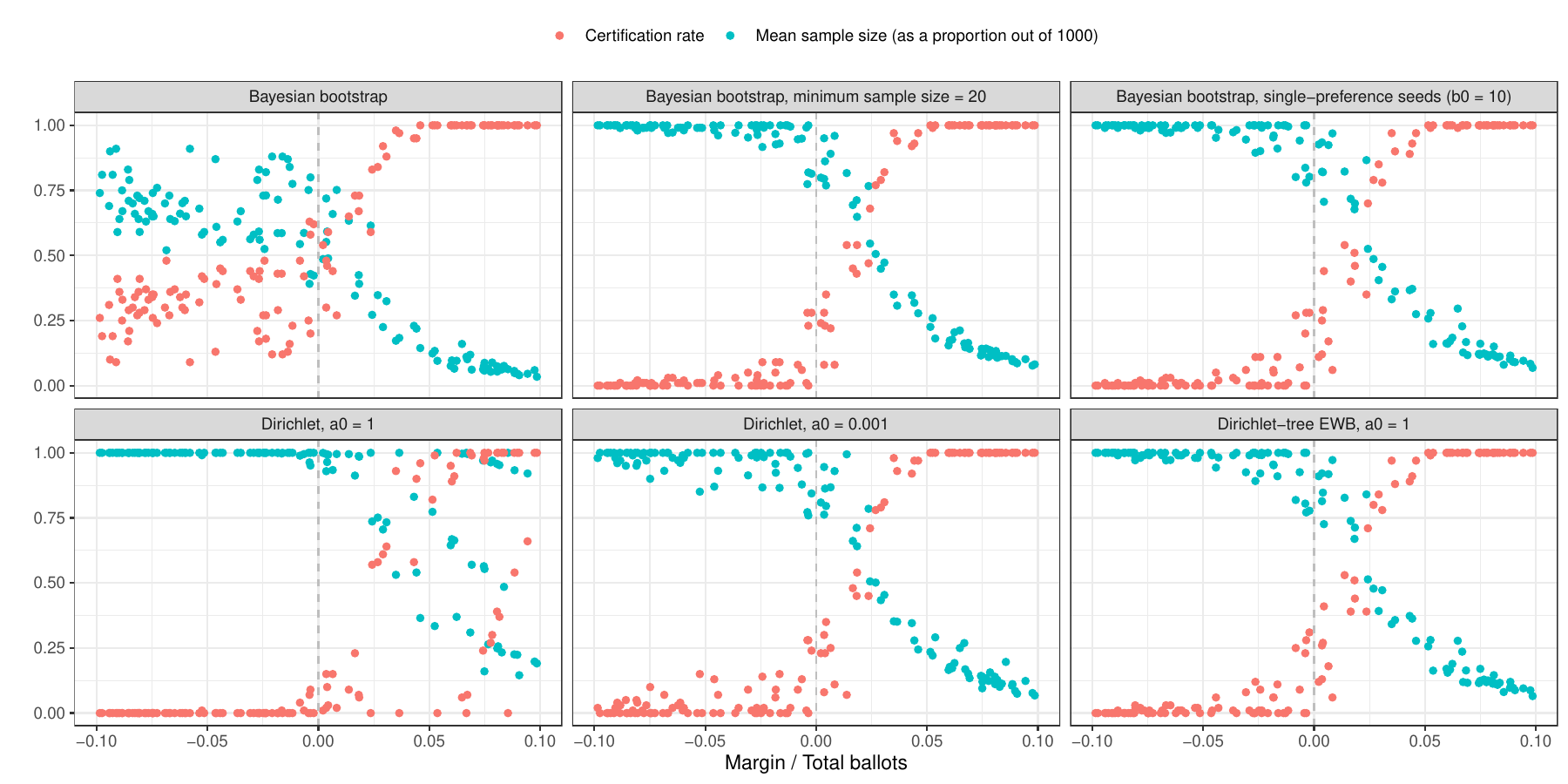}
\caption{\textbf{Performance comparison.}
Certification rate (pink; y-axis) and mean sample size (blue; y-axis, as a
proportion of the maximum allowed sample size) vs exact margin (x-axis; as a
proportion of the total ballots) for all contests where the margin for the
reported winner is between $-$10\% and 10\% of total ballots.  Each panel shows
the performance under a different prior, as labelled.}
\label{fig:certrate}
\end{figure}

\autoref{fig:certrate} summarises the performance for a large selection of
these experiments.  Each contest is represented by two points on each plot (one
for each of the two performance measures).

For most methods, performance followed the expected pattern: as the margin
increased, so did the certification rate, with a reduction in the sample size.

When the priors were made more informative, the models would typically respond
by requiring more data before certifying, thus reducing certification rates and
increasing sample sizes.  However, see \autoref{sec:posterior-bias} for some
different behaviour when using the EWB prior for some elections.

From the diverse set of contests, we can get a rough sense of what a typical
value for the risk might be for real data, by looking at the (mis)certification
rates for very close contests.  For example, for the EWB model with $a_0 = 1$,
the highest such rate is 31\%.  This would be considered too high in practice,
and could be reduced by setting a stricter threshold on the posterior
probability.  (Such settings could be explored in future.  In our current study
design we were limited to resolution of 0.01 for the posterior probabilities,
meaning that our threshold of 99\% was the highest we could set.)

The Bayesian bootstrap performed poorly when used in a default, naive way.
This was entirely due to certifying very early in the audit, before it had
enough data for its approximate posterior to stablise.  The two proposed
adaptations of the bootstrap both overcame this problem.

The modified bootstrap methods both performed similarly to the default EWB
model ($a_0 = 1$).  Any of these seem like good choices to use in practice.
The bootstrap methods are faster to compute and easier to explain, so might be
more suited for use in an actual audit.  On the other hand, the full EWB model
might be better suited for any mathematical extensions (e.g.\ as part of the
PPR; see \autoref{sec:discussion}) because it has full support over the
parameter space.

The Dirichlet with $a_0 = 1$ had poorer performance for many contests, and
completely failed for some, most likely because it is very informative (i.e.\
when $K$ is very large).  How informative it is highly depends on the number of
candidates.  This is reflected in the large variation in performance across
contests.  When $a_0 = 0.001$, the performance was more reasonable across
contests.  However, some variation is still visible.  Rather than trying to
tune a good value of $a_0$ for each contest, we suggest simply using an EWB
prior with $a_0 = 1$.

For contests with a 3-preference limit, we explored what happens if this
restriction is omitted (i.e.\ we use a mis-specified model).  Usually, this
simply made the computation slower but did not change statistical performance.
An exception is the Dirichlet priors, for which the total weight across the
branches increased substantially when the tree structure was altered.
%\footnote{%
%This was due to the way we set the weights, by starting at the leaves and
%summing up the tree.  We could account for this by setting a much smaller
%weight when using a deeper tree.}

\subsection{Pathological examples}
\label{sec:pathological}

It is not known whether there is a way to set the threshold upset probability
in Dirichlet-tree Bayesian audits to limit the risk to a pre-specified level.
While in practice they perform well, we can construct artificial contests where
they usually fail to find the correct winner.

Our contests had 10 candidates: $A$ the (true) winner, $B$ an alternate winner,
and 8 supporting candidates, $C_1, C_2, \dots, C_8$.  The ballots in the
contest were:
\begin{itemize}
\item $2000 + 2 m$ ballots of the form $[A]$,
\item $1000 - 2 m$ ballots of the form $[B]$,
\item 1000 ballots of the form $[C_i, B, A]$ for each $i$.
\end{itemize}
In the correct count of these contests, $B$ is eliminated first (and their
ballots exhaust), then each of the supporters are eliminated (and their ballots
are distributed to $A$ since $B$ is already eliminated) leading to $A$ being
the winner.
Note that $m$ is the margin in this contest.
%We can change the outcome by replacing $m$ ballots of the form $[C_i, B, A]$
%with $[B]$ for any supporting candidate $C_i$ to make $B$ win: now $C_i$ will
%be eliminated first and distribute its votes to $B$, after which the other
%supporting candidates will follow suit, ensuring that $B$ wins overall.

%The problem with this election is that
If we erroneously eliminate any of the 8 supporting candidates before $B$, then
$B$ can never be eliminated and is declared the winner. Small errors in the
count, such as would occur under random sampling, are likely to lead to at
least one of the supporting candidates being eliminated.

We used $m \in \{1, 5, 50\}$ to define a set of pathological contests.  Through
our benchmarking experiments, we evaluated how often the contests were
miscertified for candidate $B$ by the different methods.  The results are shown
in \autoref{tab:pathological}.

\begin{table}[t]
\centering
\caption{\textbf{Pathological contests.}
The (mis)certification rates for candidate $B$ when it is (incorrectly)
reported as the winner.}
\smallskip
\begin{tabular}{lrrr}
\toprule
      & \multicolumn{3}{c}{\textbf{Cert.\ rate (\%)}} \\
\cmidrule{2-4}
\textbf{Prior} &  Margin: $-1$  &  $-5$  &  $-50$     \\
\midrule
Bayesian bootstrap                                       & 84 & 86 & 76 \\
Bayesian bootstrap, minimum sample size $= 20$           & 84 & 86 & 75 \\
Bayesian bootstrap, single-preference seeds, $b_0 = 10$  & 79 & 79 & 70 \\
Dirichlet-tree EWB, $a_0 = 1$                            & 79 & 79 & 66 \\
Dirichlet, $a_0 = 0.001$                                 & 84 & 86 & 77 \\
Dirichlet, $a_0 = 1$                                     & 57 & 54 & 32 \\
\bottomrule
\end{tabular}
\label{tab:pathological}
\end{table}

Clearly there is a high chance of miscertifying any of these pathological
elections, and it is still considerable even when the margin grows.

Note that risk-limiting auditing methods will also struggle with these
pathological contests since the margins are so small.  But, rather than
incorrectly certifying the contest in favour of $B$ a large proportion of the
time, they would instead escalate to a full count of the ballots.

\subsection{Bias induced by overly informative priors}
\label{sec:posterior-bias}

We explored the effect of making the priors increasingly more informative
(e.g.\ increasing $a_0$).  The aim was mainly to help understand the behaviour
of the models; very informative priors would typically be avoided in practice.

\begin{figure}[t]
\centering
\includegraphics[width=\textwidth]{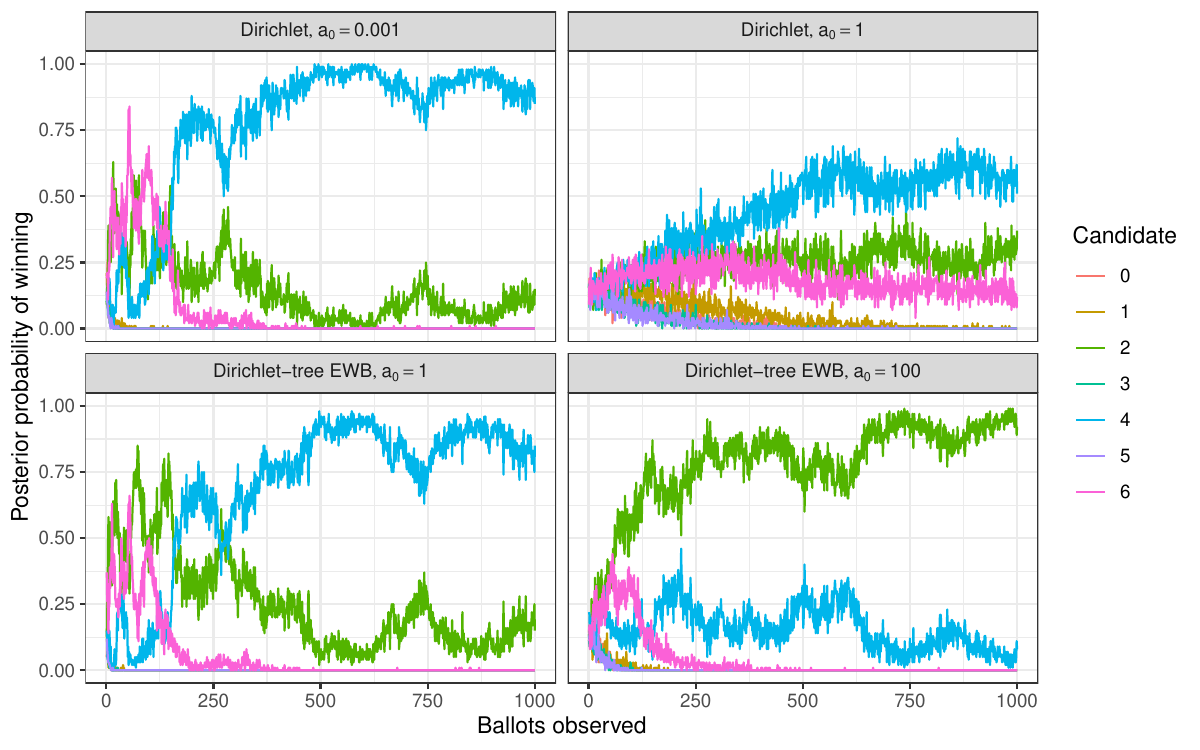}
\caption{\textbf{Posterior probabilities with weakly and strongly informative
priors.}
Similar to \autoref{fig:posterior-trace}, for a simulated audit of the Ballina
(NSW 2015) contest, using four different priors.  The true winner is candidate
4 (blue line).  The priors in the left column are weakly informative and give
similar posteriors.  The priors in the right column are strongly informative
but in different ways, resulting in very different posteriors.}
\label{fig:posterior-bias}
\end{figure}

We noticed a peculiar property of the EWB model.  For some elections, making
the prior more informative led to shifting the posterior from favouring one
candidate to strongly favouring a different candidate.  See
\autoref{fig:posterior-bias} for an example: the EWB model favours candidate 4
when $a_0 = 1$, but increasing this to $a_0 = 100$ shifts the support to
candidate 2.

This was surprising because the priors were symmetric with respect to all
candidates, and because we would normally expect that making such a prior
stronger would simply dilute any `signal' in the data.  This is indeed what
happens for most elections (not displayed here) and also happens for this
election if using a Dirichlet prior.

The reason for the surprising behaviour is that the EWB prior will provide
stronger `shrinkage' for the later preferences because the data become more
sparse as we go further down the tree.  If a candidate relies on preferences
being distributed to them in order to win, this can get disrupted if the
shrinkage is too strong for those preferences.
%For some intermediate strength of the prior, it will provide sufficient
%shrinkage to prevent the true winner from gathering enough lower-down
%preferences, but not so much shrinkage that the first-preference counts get
%completely diluted.  Such a prior will end up much more strongly favouring the
%candidate who is leading on first-preferences.
%
%The Dirichlet prior doesn't suffer from this effect because the weights in the
%branches at each level are in proportion to the data, since the prior can be
%interpreted precisely as pseudocounts of hypothetical ballots.
We only observed this biasing effect for EWB with large $a_0$, so we do not
expect this to be a problem in practice, when we would typically set $a_0
\leqslant 1$.

\subsection{Comparison with existing methods}

RAIRE (together with SHANGRLA) is an existing RLA for IRV elections, so it is
natural to ask how it compares with this new Dirichlet-tree approach.

Setting up a meaningful comparison is difficult because the methods operate
very differently.
RAIRE is an RLA, guaranteeing that miscertification rates won't exceed a given
limit.  The Bayesian Dirichlet-tree methods do not (yet) have this guarantee.
Comparing performance in terms of, for example, the mean sample size,
will not be insightful until we can calibrate the Bayesian methods.

Another difference is that RAIRE, by design, needs to know the full set of
ballots in the election in order to design its `assertions'.  It optimises its
performance under the assumption that any errors are likely to be small.  In
contrast, the Bayesian Dirichlet-tree methods operate without using this
information.  This means it will likely perform worse when the error rate is
indeed small, but it also makes it more robust and more widely applicable.

To illustrate this difference, we ran both methods with
candidate labels permuted.\footnote{%
This example is meant to be illustrative, however such a scenario is
conceivable in practice as a result of a software bug or deliberate
tampering.}
For RAIRE, this involved giving it a relabelled set of ballots to design its
assertions, but then running the audit using the real ballots.  For the
Bayesian methods, it simply involved running the audit with a different
reported winner, as described earlier (\autoref{sec:permuting-candidates}).
%For this analysis, we used one of the smaller elections and did not specify a
%limit on the sample size.  For RAIRE, we used the ALPHA martingale in
%SHANGRLA, with $d = 20$ [cite ALPHA].
The results are shown in \autoref{tab:aspen-permute}.

This scenario is one where the tabulation incorrectly interpreted \emph{every}
ballot, a challenging case for RAIRE, because it is likely to make its
assertions false even when the reported winner still won.  For nearly all
permutations, it responds with the need to do a full count of the ballots.  The
Bayesian methods only do this if the true winner gets relabelled, but not
otherwise.

Whether or not this behaviour is desireable will depend on the goals of your
audit.  What is clear from this example, at least, is that the two methods can
differ substantially.  In practice, both methods could be run in tandem with
the same set of data, each method providing its own benefits.

\begin{table}
\centering
\caption{\textbf{Comparison of methods after permuting candidate labels.}
Contest: Aspen Mayoral election 2009 (5 candidates, 2544 ballots).  The true
winner was candidate 4.  Methods: RAIRE/SHANGRLA, $d = 10$; Dirichlet-tree EWB,
$a_0 = 1$.  For each of the 120 permutations, and for each method, we show the
certification rate for the reported winner and the mean sample size to audit
the contest (with no sample size limit imposed).  A sample size of 2544 is a
full count of all ballots.  Each row represents a group of permutations that
gave rise to the same performance.  The first permutation is the identity (no
changes to the candidate labels).  All other permutations involve some
candidates being relabelled.}
\smallskip
\begin{tabular}{llrrrr}
\toprule
\multicolumn{2}{l}{\textbf{Permutations}}     &
\multicolumn{2}{c}{\textbf{Cert.\ rate (\%)}} &
\multicolumn{2}{c}{\phantom{i}
                   \textbf{Mean samp.\ size}} \\
\cmidrule(r){1-2}
\cmidrule{3-4}
\cmidrule(l){5-6}
\# permutations \phantom{i} &
Reported winner \phantom{i} &
RAIRE & EWB & RAIRE & EWB \\
\midrule
 1 (no change) & 4 & 100 & 100 &  845 &  335 \\
 1             & 4 & 100 & 100 &  855 &  335 \\
22             & 4 &   0 & 100 & 2544 &  335 \\
24             & 1 &   0 &   1 & 2544 & 2520 \\
72     & 2, 3 or 5 &   0 &   0 & 2544 & 2544 \\
\bottomrule
\end{tabular}
\label{tab:aspen-permute}
\end{table}

% ---------------------------------------------------------------------------

\section{Discussion}
\label{sec:discussion}

We have implemented and provided a thorough analysis of a method for
ballot-polling Bayesian audits of IRV elections.  The method is computationally
efficient, even for elections with a large number of candidates or ballots.

The flexibility of our method allows straightforward extension to STV and other
ranked voting scenarios, by changing the social choice function (to cater for
different voting systems) and adapting the tree structure (to cater for any
restrictions on the design of the ballots, or to improve efficiency).  For
example, for Australian Senate elections (which use STV), the first split in
the tree could distinguish between an `above the line' vote versus a `below the
line' vote (a detail specific to Australian Senate ballots that impacts how
they are counted).

One limitation of our approach is that it is not yet known whether there is an
easy way to compute or impose a risk limit.  We previously mentioned some ideas
to tackle this \cite{everest2022evote}.  We flesh them out here in more detail:
\begin{itemize}
\item We could determine the maximum risk by deriving the worst-case
configuration of true ballots, as was previously done for 2-candidate contests
\cite{huang_unified_2020}.  If successful, this would give either a formula or
an efficient algorithm to determine a threshold for the posterior probability,
given a desired risk limit.

\item We could derive a prior-posterior ratio (PPR) martingale
\cite{waudby-smith-2020-confseq-wor} using the Dirichlet-tree model.  This
would almost directly give a risk-limiting method, with the main challenge
being to devise an efficient algorithm to determine whether the PPR threshold
is achieved given a sample of ballots.  While such a method would use a
Dirichlet-tree model, it would not actually be a Bayesian audit: the stopping
criterion is not based on a posterior threshold.
\end{itemize}
%
%A separate challenge, which is important in practice, is to go beyond
%ballot-polling audits.  For example, for elections where cast vote records
%(CVRs) are available, a comparison audit would be more resource-efficient.

% ---------------------------------------------------------------------------

\subsubsection{Acknowledgements.}

We thank Ronald Rivest for many helpful suggestions for improving the paper.
This work was supported by the University of Melbourne's Research Computing
Services and the Petascale Campus Initiative; and by the Australian Research
Council (Discovery Project DP220101012).

% ---------------------------------------------------------------------------

\bibliographystyle{splncs04}
\bibliography{references}

% ---------------------------------------------------------------------------

\end{document}